\documentclass[10pt, letterpaper, conference]{IEEEtran}
\IEEEoverridecommandlockouts
\usepackage{booktabs} 
\usepackage{makecell}
\usepackage{flushend}

\usepackage{cite}
\usepackage{amsmath,amssymb,amsfonts}
\usepackage{algorithmic}
\usepackage{graphicx}
\usepackage{textcomp}
\usepackage{xcolor}
\def\BibTeX{{\rm B\kern-.05em{\sc i\kern-.025em b}\kern-.08em
    T\kern-.1667em\lower.7ex\hbox{E}\kern-.125emX}}
\begin{document}

\title{Accelerating QUIC's Connection Establishment\\ on High-Latency Access Networks}

\author{\IEEEauthorblockN{Erik Sy, Tobias Mueller, Moritz Moennich and Hannes Federrath}
\IEEEauthorblockA{University of Hamburg\\
Hamburg, Germany}
}

\maketitle

\begin{abstract}
A significant amount of connection establishments on the web require a prior domain name resolution by the client.
Especially on high-latency access networks, these DNS lookups cause a significant delay on the client's connection establishment with a server.
To reduce the overhead of QUIC's connection establishment with prior DNS lookup on these networks, we propose a novel QuicSocks proxy.
Basically, the client delegates the domain name resolution towards the QuicSocks proxy.
Our results indicate, that colocating our proxy with real-world ISP-provided DNS resolvers provides great performance gains.
For example, 10\% of our 474 sample nodes distributed across ISP's in Germany would save at least 30~ms per QUIC connection establishment.
The design of our proposal aims to be readily deployable on the Internet by avoiding IP address spoofing, anticipating Network Address Translators and using the standard DNS and QUIC protocols.
In summary, our proposal fosters a faster establishment of QUIC connections for clients on high-latency access networks.
\end{abstract}

\begin{IEEEkeywords}
QUIC Transport Protocol, SOCKS Proxy, DNS, QuicSocks Proxy
\end{IEEEkeywords}

\section{Introduction}

For U.S. households latency is the main web performance bottleneck for broadband access networks exceeding a throughput of 16~Mbit/sec~\cite{sundaresan2013community}.
Depending on the user's location and the deployed access technology like cable, fiber, Digital Subscriber Line (DSL), Long-Term Evolution (LTE), or satellite, users may experience significant network latencies.
High-latency links reduce the user's quality of experience during web browsing~\cite{Varvello:2016:EPC:2999572.2999590} and negatively impact the per-user revenue of online service provider~\cite{velocity}.
Thus, optimizing the web performance on such existing high-latency network links is an important task.
In this paper, we focus on improving the time to first byte which contributes up to 21\% of the page load time for popular websites~\cite{sundaresan2013community}.
In detail, we improve the delay of QUIC's connection establishment with prior DNS lookup on high-latency links.
QUIC replaces the TLS over TCP protocol stack within the upcoming HTTP/3 version~\cite{ietf-quic-http-20}.
As the web is built upon the Hypertext Transfer Protocol (HTTP) and the standardization of QUIC receives widespread support, the QUIC protocol is expected to be widely deployed on the Internet in the forthcoming years.

Our proposal assumes, that Internet Service Providers (ISP) aim to improve their clients' quality of experience during web browsing.
This assumption is substantiated by ISPs providing recursive DNS resolvers to accelerate their client's DNS lookups.
In this work, we propose ISP-provided proxies to reduce the delay of their client's QUIC connection establishments.
Instead of conducting a DNS lookup and waiting for the response, this design allows the client to directly send its initial QUIC messages to the novel QuicSocks proxy.
Upon receiving these messages, the proxy resolves the domain name and forwards the messages to the respective QUIC server.
After the end-to-end encrypted connection between the client and the server is established, the connection is seamlessly migrated to the direct path between these peers.  

These novel QuicSocks proxies can accelerate the client's connection establishment to a server, if they perform faster DNS lookups and/or have a lower network latency to the QUIC server compared to the client.
A favorable network topology would place a QuicSocks proxy colocated with the ISP-provided DNS resolver in an on-path position between the client and the server.

Our proposal can be applied to many high-latency links.
Within the next years, enterprises like SpaceX, OneWeb, and Telesat plan to launch thousands of satellites for global broadband connectivity aiming to  provide Internet access to millions of people~\cite{forbes}.
This presents a well-suited application area for our proposal because of the significant latencies of about 30\,ms between the client and the ISP's ground station~\cite{dslreports}. 

In summary, this paper makes the following contributions:

\begin{itemize}

\item We propose the novel QuicSocks design that allows clients to send initial handshake messages without a prior resolution of the domain name. The name resolution is conducted by a QuicSocks proxy from a more favorable position in the ISP's network to accelerate the connection establishment.

\item We evaluate our proposal by assuming a colocation of the ISP-provided DNS resolver with the QuicSocks proxy. Results based on our analytical model indicate for a QUIC connection establishment accelerations between 33\% and 40\% on a U.S. mobile LTE network. Furthermore, our measurements of real-world network topologies indicate the feasibility of significant performance gains for clients on high-latency access networks. For example. 10\% of the investigated clients save at least 30~ms to complete their QUIC handshake.

\item We implemented a prototype of our proposal to demonstrate its real-world feasibility. Our results indicate, that the computations of the QuicSocks proxy itself are lightweight and contribute less than 1.2~ms to a QUIC connection establishment.

\end{itemize}

The remainder of this paper is structured as follows: Section~\ref{sec:Problem} introduces the QUIC and the SOCKS protocol and describes the performance problem that we aim to solve. 
Section~\ref{sec:Design} summarizes the proposed QuicSocks design and evaluation results are presented in Section~\ref{sec:Evaluation}.
Related work is reviewed in Section~\ref{sec:Related}, and Section~\ref{sec:Conclusion} concludes the paper.

\section{Background and problem statement}\label{sec:Problem}

In this section, we first describe the QUIC protocol which is deployed in HTTP version 3. 
Subsequently, we review the SOCKS protocol used to exchange network packets between a client and a server through an intermediate proxy.
Then, we investigate the performance problem of DNS queries on subsequent connection establishments.

\subsection{QUIC transport protocol}

In this paper, we refer to QUIC's draft version~20 of the Internet Engineering Task Force (IETF) as the QUIC protocol~\cite{ietf-quic-transport-20}.
The QUIC transport protocol will replace TLS over TCP in the upcoming HTTP/3 network protocol~\cite{ietf-quic-http-20}, which is closely tied to the world wide web.
Thus, deployment of HTTP/3 on the web will significantly contribute to QUIC's adoption on the Internet in the forthcoming years.
Compared to TLS over TCP, the UDP-based QUIC protocol allows for faster connection establishments~\cite{DBLP:journals/corr/abs-1903-09466}, mitigates head-of-line blocking~\cite{langley2017quic}, and can be extended because of a lower interference through middleboxes~\cite{honda2011still}.

In the following, we provide details on two mechanisms of the QUIC protocol that our proposed QuicSocks approach makes use of.
These are a challenge-response mechanism in QUIC's connection establishment known as stateless retry and QUIC's connection migration that allows transferring an established connection to a new endpoint address.

\paragraph{Stateless retry}

\begin{figure}[t]
\centering
\includegraphics[width=0.47 \textwidth]{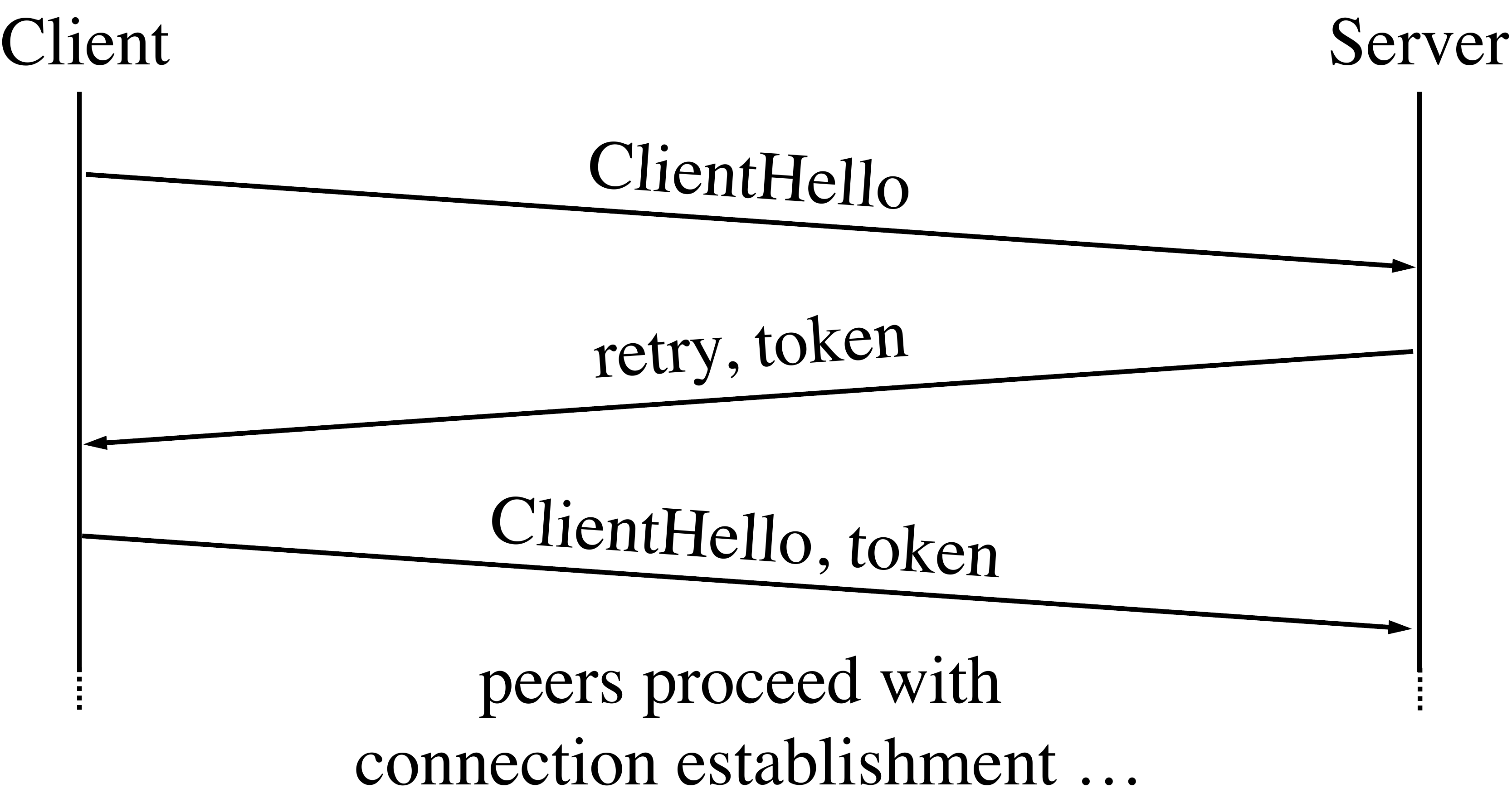}
\caption{Schematic of QUIC's stateless retry mechanism.}
\label{fig:Quic_retry}
\end{figure}

The stateless retry mechanism can be optionally used by QUIC servers to validate the source address claimed by a client before proceeding with the cryptographic connection establishment.
As shown in Figure~\ref{fig:Quic_retry}, the server responds to the client's initial connection request with a retry message that contains a source address token.
This token is opaque to the client and contains information about the client's source address.
Subsequently, the client returns this token together with its previously sent ClientHello message.
Upon receiving this message from the client, the server first validates the presented token.
A token is valid for a connection request if the client's claimed source address matches the address encoded in the token.
In this case, the server assumes that the client can receive packets at the claimed address and proceeds with the cryptographic connection establishment.
A stateless retry presents a performance limitation as it adds a round-trip time to the connection establishment.
However, it supports QUIC servers to protect against denial-of-service attacks.
Therefore, QUIC servers are likely to use these optional stateless retries when experiencing many connection requests from source addresses with unresponsive clients.

\paragraph{Connection migration}

QUIC connections are associated with connection IDs that allow the identification of a connection independently of the used source address and port number.
Connection IDs allow connections to survive an endpoint's change of the IP address and/or port number which might occur because of NAT timeouts and rebinding~\cite{hatonen2010experimental}, or clients changing their network connectivity.
Only QUIC clients can initiate connection migrations to a different endpoint's IP address and/or port number.
However, the client must wait until the handshake is completed and forward secure keys are established before initiating the connection migration.
The endpoints might use multiple network paths simultaneously during the connection migration.
The peers can optionally probe a new path for peer reachability before migrating a connection to it.
However, when a connection is migrated to a new path the server must ensure the client is reachable via this path before sending large amounts of data.
Furthermore, the peers need adapting their sending rate to the new path by resetting their congestion controller and round-trip time estimator.

\subsection{SOCKS protocol}

\begin{figure}[t]
\centering
\includegraphics[width=0.47 \textwidth]{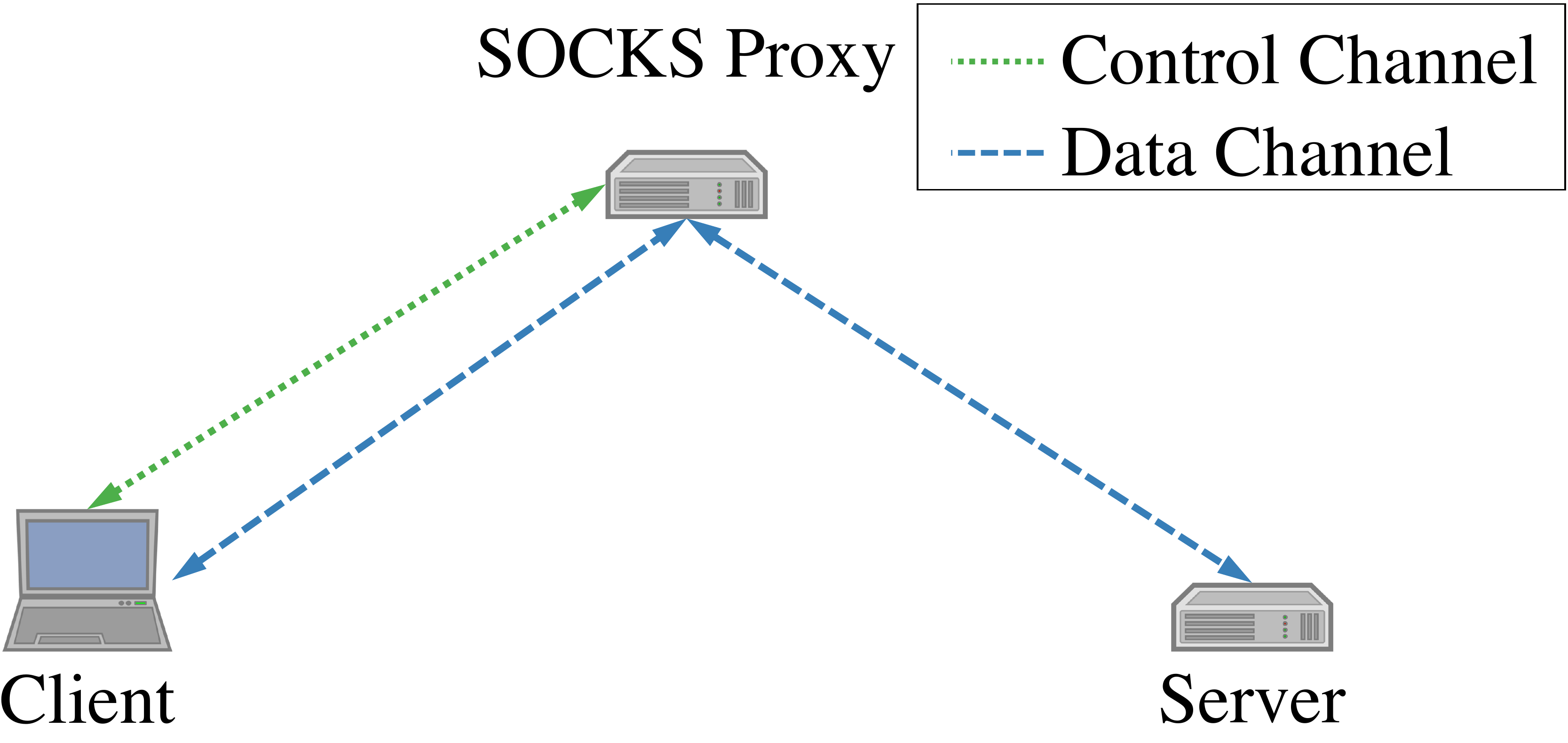}
\caption{Schematic of a network packet exchange between client and server employing an intermediate SOCKS proxy.}
\label{fig:SOCKS}
\end{figure}

RFC 1928 describes the current version of the SOCKS protocol~\cite{rfc1928}.
Usages of SOCKS proxies include the traversal of network firewalls~\cite{rfc1928}, the translation between IPv6 and IPv4 address space~\cite{rfc3089}, and privacy-enhancing technologies such as TOR onion routing~\cite{dingledine2004tor}.
Figure~\ref{fig:SOCKS} provides a schematic of a connection between client and server through a SOCKS proxy.
To begin with, the client establishes a TCP connection to the proxy's port 1080.
This connection is used by the client and the SOCKS proxy to exchange control messages.
For example, the client can use this control channel for authentication or to request a new connection to a server.
The SOCKS protocol supports the exchange of UDP datagrams between the client and server.
As a result, QUIC connections can be established via default SOCKS proxies.
For this purpose, the client sends a UDP associate request to the socks proxy.
If the client's request is approved, the SOCKS proxy responses with a source address and port number to which the client can send the UDP datagrams to be relayed to the server.
Subsequently, the client attaches a SOCKS request header to its UDP datagrams and sends them to the indicated port number/ IP address.
Upon receiving these UDP datagrams, the proxy will remove the request header and send them from its own source address to the server.
The server will send its response to the proxy server, which will then relay it to the client.
Note, that the SOCKS protocol allows clients to delegate the task of DNS name resolution.
For this purpose, the client includes the domain name within its request header.
Subsequently, the SOCKS proxy resolves this domain name and relays the packets to the corresponding destination.

\subsection{Delays caused by high latencies to recursive DNS resolvers}

The Domain Name System (DNS) is responsible for resolving domain names into IP addresses.
Many operating systems or web browsers have a local DNS cache.
However, between 12.9\% and 20.4\% of a user's established TCP connections directly follow a DNS query~\cite{jung2002dns}.
A popular website requires connections to about 20 different hostnames~\cite{DBLP:journals/corr/abs-1902-02531}.
Hence, the user conducts on average between 2.6 and 4.1 fresh DNS queries per website retrieval.
Each of these DNS queries delays the subsequent connection establishment to the server serving the queried hostname.
Furthermore, websites usually have a nested hierarchy of requests to different hostnames~\cite{DBLP:journals/corr/abs-1902-02531}.
If such nested requests to different hostnames require each a DNS query by the client, then the website loading is delayed by the time required for these sequential DNS queries.

\begin{figure}[t]
\centering
\includegraphics[width=0.47 \textwidth]{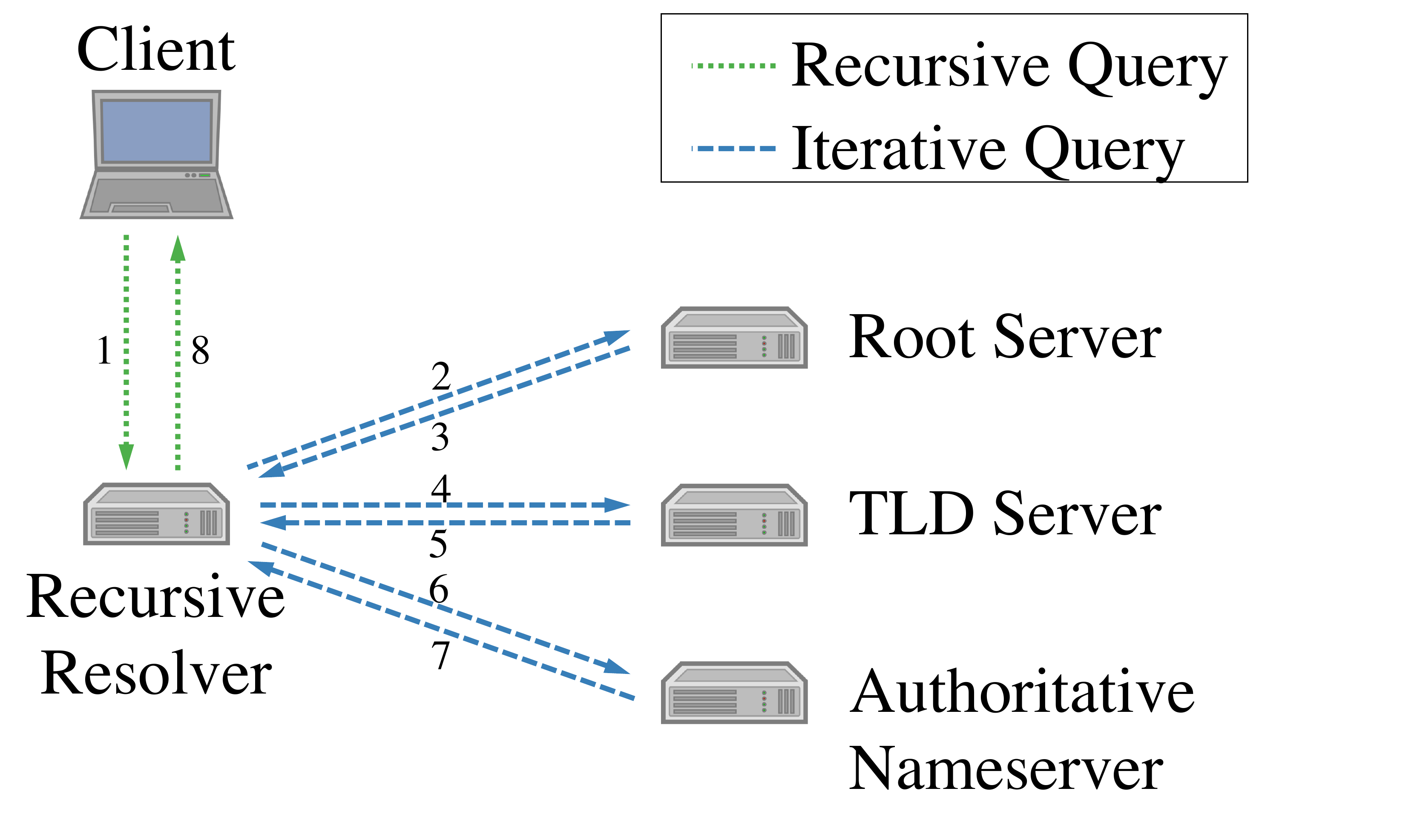}
\caption{Schematic of a complete DNS query using a recursive resolver.}
\label{fig:Recursive_DNS}
\end{figure}

In this paper, we assume that the client can resolve either a domain name using its local cache or needs to query recursively its DNS resolver as shown in Figure~\ref{fig:Recursive_DNS}.
If the recursive resolver has a cache miss for the queried domain name, it starts an iterative query.
The arrows two to seven in Figure~\ref{fig:Recursive_DNS} indicate such a complete iterative query involving the DNS root server, Top Level Domain (TLS) server, and finally the authoritative nameserver of the respective domain name.
DNS recursive resolvers make extensive use of caching with reported hit rates larger than 80\%~\cite{jung2002dns}.
Thus, the round-trip time (RTT) between the client and the recursive resolver can present a significant source of delay for a DNS query.
Studies of home routers in the US indicate typical RTT between 5\,ms and 15\,ms to their ISP-provided DNS resolver~\cite{sundaresan2013community}.
However, a fraction of about 5\% of the users experience a RTT longer than 20\,ms~\cite{zhu2015connection}.
Studies with the popular third-party resolver Google Public DNS indicate a median RTT of 23\,ms, while between 10\% and 25\% of the measurement nodes experienced RRTs longer than 50\,ms~\cite{zhu2015connection}.
For users having a downstream throughput of more than 16~Mbits/sec, the page load time highly depends on their network latency and DNS query time compared to their available throughput~\cite{sundaresan2013community}.
As a result, especially clients having a high network latency to their resolver require technological improvements to reduce their experienced DNS query time to achieve faster website retrievals.

\section{QuicSocks}\label{sec:Design}

In this section, we introduce the QuicSocks design.
This novel approach improves the latency of QUIC's connection establishments that directly follow a DNS lookup.
First, we summarize our design goals, before we present QuicSocks.
Finally, we describe the implementation of our QuicSocks prototype.

\subsection{Design goals}

We aim to develop a solution that supports the following goals:
\begin{enumerate}
\item Deployable on today's Internet which excludes approaches requiring changes to middle-boxes, kernels of client machines, the DNS protocol, or the QUIC protocol. 
\item Reduces the latency of QUIC's connection establishments that require a prior DNS lookup.
\item Does not make use of IP address spoofing as this practice conflicts with RFC~2827~\cite{rfc2827}.
\item Supports clients behind Network Address Translators (NAT).
\item Guarantees confidentiality by assuring end-to-end encryption between the client and the web server.
\item Limits the consumption of the proxy's bandwidth.
\item Privacy assurances similar to using a recursive DNS resolver.
\end{enumerate}

\subsection{Design}

In this section, we present the protocol flow of a connection establishment using a QuicSocks proxy.

First, the client needs to establish a control channel with the QuicSocks proxy.
The client learns via the control channel which port of the QuicSocks proxy can be used for its connection establishments.
A single control channel can be used to establish several QUIC connections via the proxy.
Furthermore, the control channel is used by the proxy to validate the client's claimed source address.

\begin{figure}[t]
\centering
\includegraphics[width=0.47 \textwidth]{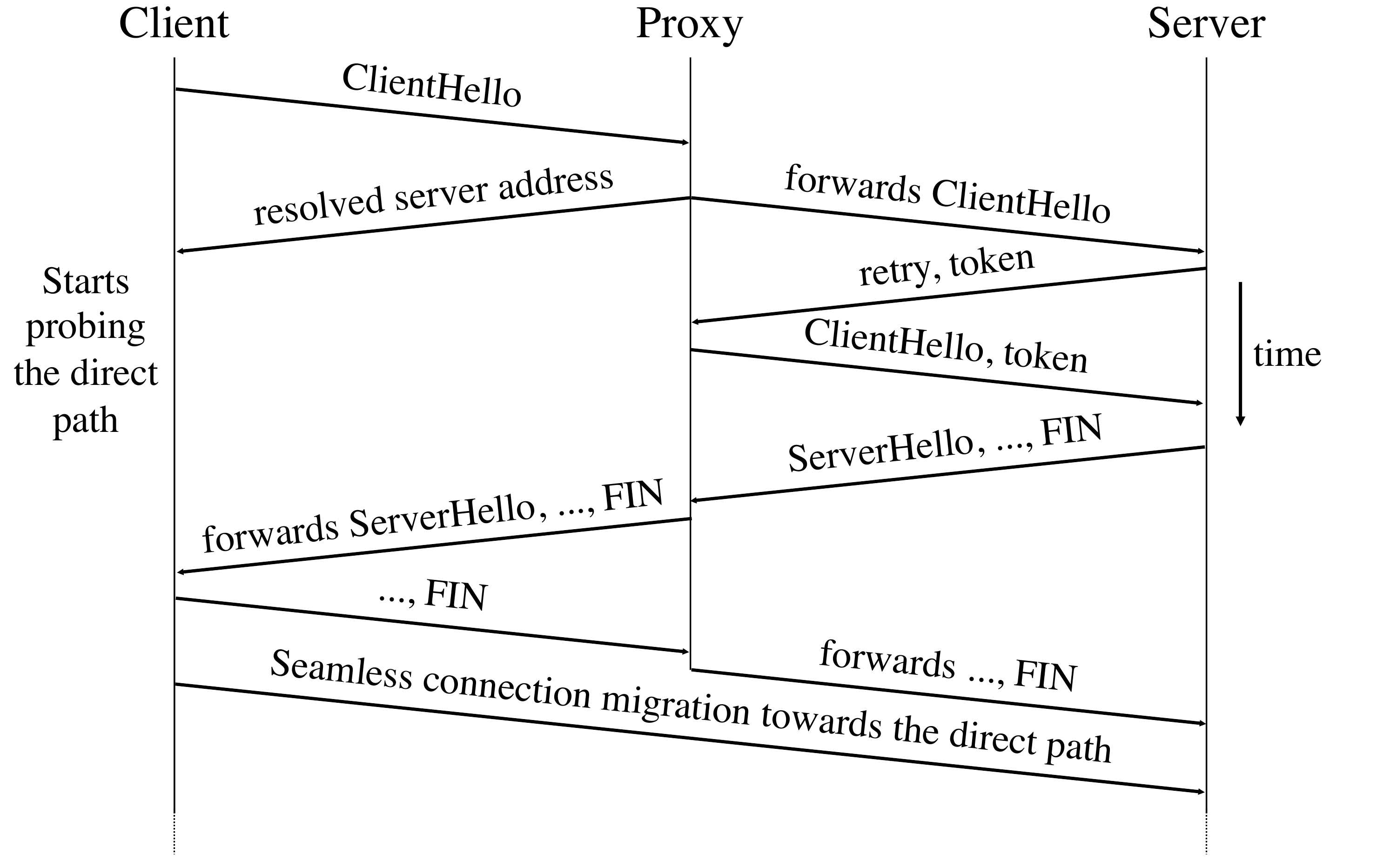}
\caption{Protocol flow of a QUIC handshake via the proposed QuicSocks proxy, who resolves the server's IP address. The handshake includes an optional stateless retry initiated by the server. After the server and client established forward secure keys and exchanged FIN messages, the client initiates the connection migration towards the direct path.}
\label{fig:QuicSocks_Flow}
\end{figure}

Subsequently, the establishment of a single QUIC connection follows the protocol flow shown in Figure~\ref{fig:QuicSocks_Flow}.
Note, that each UDP datagram exchanged between the client and server is encapsulated and carries a request header as common in the SOCKS protocol~\cite{rfc1928}.
To begin the connection establishment, the client sends its QUIC ClientHello message to the QuicSocks proxy indicating the domain name of the destination server in the SOCKS' request header.
Upon receiving this message, the proxy authenticates the client based on the datagrams' encapsulation and caches this message.
Subsequently, the proxy does a DNS lookup for the presented domain name and forwards the ClientHello message to the destination's server IP address.
Next, the proxy forwards also the obtained DNS response to the client.
Note, that the QuicSocks proxy sends all forwarded datagrams from its own source address.
Upon receiving the DNS response from the proxy, the client starts probing the direct path to the respective web server to prepare a seamless connection migration to this new path.

Upon receiving the forwarded ClientHello, the server can optionally conduct a stateless retry as shown in Figure~\ref{fig:QuicSocks_Flow}.
In this case, the server returns a retry message and an address validation token to the proxy.
Receiving such a request for a stateless retry, the proxy resends the cached ClientHello message and along with the received address validation token.
This challenge-response mechanism allows the QUIC server to validate the claimed source address before proceeding with the cryptographic connection establishment.
Following the default QUIC handshake, the server proceeds by sending messages including the ServerHello and the FIN, which signals that the server established forward-secure encryption keys.
Receiving these messages of the cryptographic connection establishment, the proxy forwards them to the client.
Based on these messages, the client validates the server's identity and computes its forward-secure encryption keys.
To complete the handshake, the client sends its FIN message via the proxy to the server.
Subsequently, the client migrates the established connection towards the direct path between client and server.
The connection migration reduces the system utilization of the QuicSocks proxy and possibly leads to shorter round-trip times between client and server.

\subsection{Implementation}

The implementation of our proposal aims to demonstrate its real-world feasibility.
Our implemented prototype is capable of establishing a connection via the default SOCKS protocol and subsequently migrate the connection to the direct path between QUIC server and client.
Our modified client is written in about 350 lines of Rust code and make use of the rust-socks (v0.3.2) and quiche (v0.1.0-alpha3) libraries.
Rust-socks provides an abstraction of a SOCKS connection with
an interface that is similar to the operating system's UDP sockets and allows to transparently use a SOCKS connection.
Quiche is an experimental QUIC implementation that separates protocol messages from socket operations which accommodates our use-case of switching between SOCKS sockets and the operating system's UDP sockets within the same QUIC connection.
In detail, we modified quiche's example client implementation to perform a QUIC handshake through a rust-socks socket.
Once the connection establishment is completed, we switch to a new operating system UDP socket to communicate with the QUIC server over the direct path.
Note, that the server's IP address required to conduct this switching is provided by the datagram header of the default SOCKS protocol.

Furthermore, we adapted our client implementation to measure the required time for a connection establishment.
The time is measured from the request to establish a connection until the QUIC handshake is completed.
In total, we implemented these time measurements for three different connection situations.
The first connection situation includes additionally the overhead required to establish the connection with the SOCKS proxy.
The second connection situation assumes an established SOCKS connection and measures only the time required to establish a QUIC handshake employing a SOCKS proxy.
Finally, the last connection situation conducts a time measurement for a plain QUIC connection establishment without using a SOCKS proxy.

Note, that our prototype does not provide a complete QuicSocks implementation because we did not apply changes to the used SOCKS proxy.
As a result, the used proxy does not support the stateless retry mechanism as proposed.
Furthermore, our proxy does not provide the client with the resolved QUIC server address directly after the DNS lookup.
Instead, within our test setup, the client retrieves the QUIC server address from the SOCKS encapsulation of the forwarded server response.
Hence, our client implementation does not start validating the direct path between client and server before migrating the connection.

\section{Evaluation}\label{sec:Evaluation}
In this section, we evaluate the proposed connection establishment via QuicSocks proxies.
To begin with, we investigate feasible performance improvements of our proposal compared to the status quo via an analytical model.
Then, we conduct latency measurements between clients, servers, and DNS resolvers to approximate real-world delays for QuicSocks proxies that are colocated with the respective DNS resolver.
Finally, we present performance measurements using our QuicSocks prototype.

\subsection{Analytical evaluation}

\begin{figure}[t]
\centering
\includegraphics[width=0.47 \textwidth]{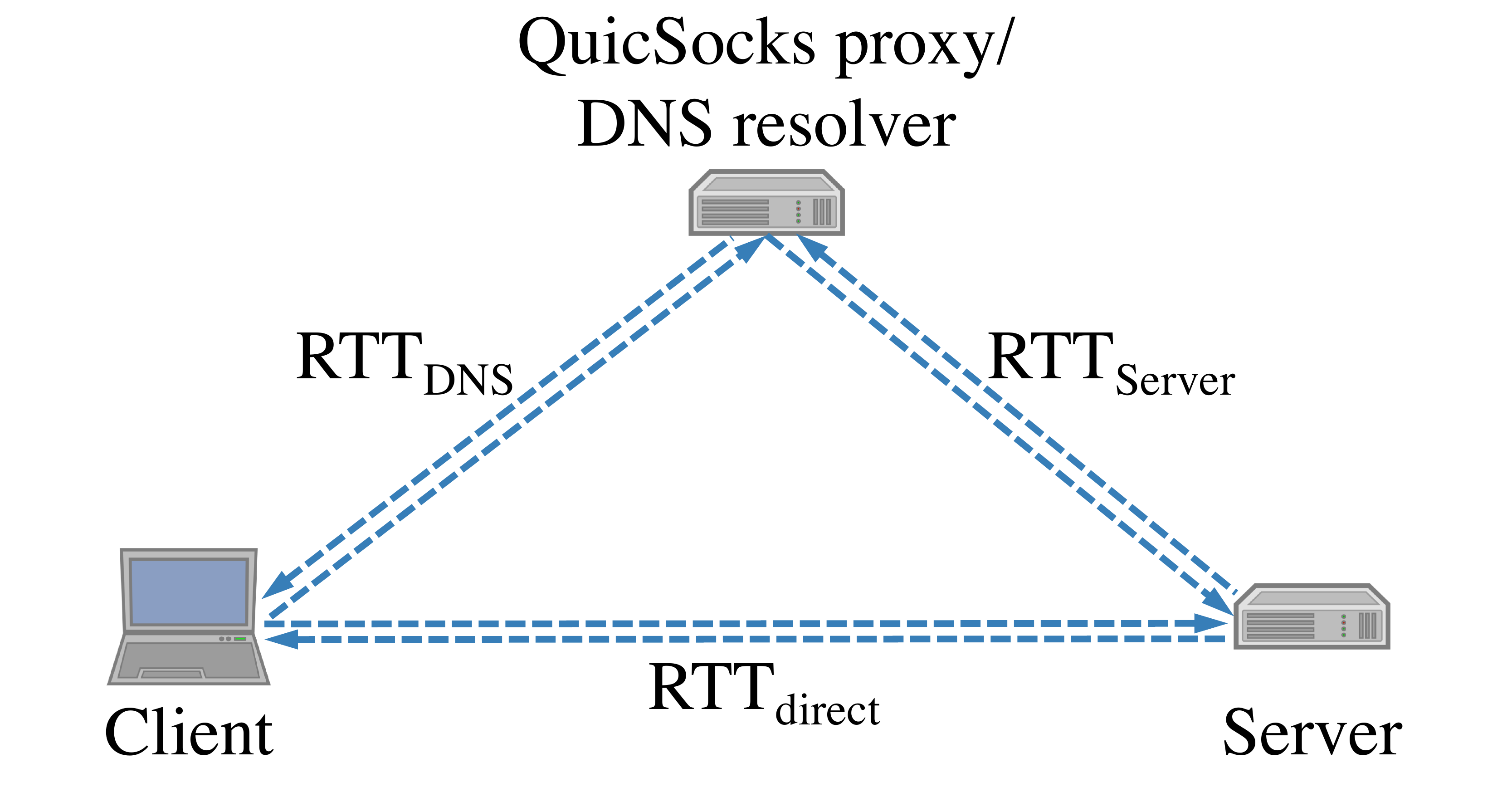}
\caption{Network topology with colocation of DNS resolver and QuicSocks proxy. Furthermore, this overview marks the round-trip time (RTT) between different peers. }
\label{fig:QuicSocks_Setup}
\end{figure}

The performance benefit of employing a QuicSocks proxy for the connection establishment depends on the network topology.
For reasons of clarity, we assume in our analytical model a colocation of the DNS resolver and the QuicSocks proxy (see Figure~\ref{fig:QuicSocks_Setup}).
Furthermore, our model is reduced to the network latency between the involved peers.
As shown in Figure~\ref{fig:QuicSocks_Setup}, we denote the round-trip time between client and DNS resolver/ QuicSocks proxy as $\text{RTT}_{\text{DNS}}$.
$\text{RTT}_{\text{direct}}$ and $\text{RTT}_{\text{Server}}$ describe the round-trip time between server and client, and between server and QuicSocks proxy, respectively.  
 
 \begin{table}[htbp]
   \caption{Comparison of the required network latency to resolve a domain name and establish a QUIC connection using the status quo and the proposed QuicSocks proxy.}
  \label{tab:analytical_model}
 \centering
 \begin{tabular}{lrr}
\toprule
\multicolumn{1}{c}{Stateless} & \multicolumn{2}{c}{Latency to establish the connection (incl. DNS)} \\
  \cmidrule(){2-3}
\multicolumn{1}{c}{retry} & \multicolumn{1}{c}{Status quo} & \multicolumn{1}{c}{Proposal}\\
 \midrule
w/o &$\text{RTT}_{\text{DNS}} + \text{RTT}_{\text{direct}}$ & $ \text{RTT}_{\text{DNS}} +\text{RTT}_{\text{Server}}$\\
 with &$ \text{RTT}_{\text{DNS}} + 2*\text{RTT}_{\text{direct}}$ & $\text{RTT}_{\text{DNS}} + 2* \text{RTT}_{\text{Server}}$\\
 \bottomrule
 \end{tabular}
 \end{table}
 Table~\ref{tab:analytical_model} presents the evaluation results for our analytical model.
 A connection without stateless retry requires $\text{RTT}_{\text{DNS}}$ to resolve the domain name and subsequently $\text{RTT}_{\text{direct}}$ to establish the connection between client and QUIC server using the status quo.
However, to establish the same connection via a QuicSocks proxy the sum of $\text{RTT}_{\text{DNS}}$ and $\text{RTT}_{\text{Server}}$ is required.
Note, that we define a connection to be established when the client and the server computed their forward-secure encryption keys and are ready to send application data.
Thus, we may count a connection as established before the client's FIN message has been processed by the server.
With respect to stateless retries, we observe that the delay of the connection establishments increase for the status quo and our proposal by a $\text{RTT}_{\text{direct}}$ and a $\text{RTT}_{\text{Server}}$, respectively.
In total, our analytical model indicates our proposal outperforms the current status quo if $\text{RTT}_{\text{Server}}$ is smaller than $\text{RTT}_{\text{direct}}$.
In this case, we find that the reduced delay of the connection establishment without stateless retry is equal to the difference between $\text{RTT}_{\text{Server}}$ and $\text{RTT}_{\text{direct}}$.
Moreover, the benefit of our proposal is doubled if the connection establishment requires a stateless retry.

Our proposal achieves its worst performance when the client is colocated with the server.
However, the best performance can be realized when the DNS resolver, the QuicSocks proxy, and the server are colocated.

In the following, we assume an ISP provides a DNS resolver/ QuicSocks proxy half-way, on-path between client and server.
Note, that the client's latency to the first IP hop (last mile latency) contributes between 40\% and 80\% of a typical $\text{RTT}_{\text{direct}}$~\cite{Sundaresan:2011:BIP:2018436.2018452}.
A typical $\text{RTT}_{\text{direct}}$ in the LTE mobile network of the U.S. is 60\,ms to reach popular online services~\cite{opensignal}.
For this example, we assume  $\text{RTT}_{\text{DNS}}$ and $\text{RTT}_{\text{Server}}$ to be each 30\,ms, while $\text{RTT}_{\text{direct}}$ is 60\,ms.

 \begin{table}[htbp]
   \caption{Comparison of the required network latency to resolve a domain name and establish a QUIC connection using the status quo and the proposed QuicSocks proxy via an average U.S. mobile LTE connection.}
  \label{tab:analytical_model_results}
 \centering
 \begin{tabular}{lrr}
\toprule
\multicolumn{1}{c}{Stateless} & \multicolumn{2}{c}{Latency to establish connection (incl. DNS)} \\
  \cmidrule(){2-3}
\multicolumn{1}{c}{retry} & \multicolumn{1}{c}{Status quo} & \multicolumn{1}{c}{Proposal}\\
 \midrule
w/o & 90\,ms & 60\,ms\\
 with &150\,ms & 90\,ms\\
 \bottomrule
 \end{tabular}
 \end{table}

Table~\ref{tab:analytical_model_results} provides the results for this example.
We find, that our proposal accelerates the connection establishment by 30\,ms and 60\,ms depending on the requirement of a stateless retry.
In summary, the total delay overhead of the connection establishment is reduced by up to 40\% and at least 33.3\%.
Note, that the absolute benefit of our proposal is even higher for 3G networks where $\text{RTT}_{\text{direct}}$ in the U.S. is on average between 86\,ms and 137\,ms depending on the mobile network provider~\cite{opensignal}. 

\subsection{Real-world network topologies}

Our analytical evaluation indicates, that our proposal can significantly reduce the latency of a QUIC connection establishments with a prior DNS query if the QuicSocks proxy has a favorable position in the network topology.
In this section, we investigate real-world network topologies to approximate the feasible performance benefit of QuicSocks proxies when they are colocated with ISP-provided DNS resolvers.
We begin by describing our applied methodology to measure real-world network topologies.
Subsequently, we evaluate the QuicSocks proposal based on our collected data.

\paragraph{Data collection}

Our data collection aims to measure  $\text{RTT}_{\text{DNS}}$, $\text{RTT}_{\text{Server}}$, and $\text{RTT}_{\text{direct}}$ for different real-world clients.
We use nodes of the RIPE Atlas network~\cite{ripe} to represent our clients.
These RIPE Atlas nodes allow us to conduct custom ping measurements and DNS queries.
The selected nodes are in different autonomous systems all over Germany including home networks and data centers.
Furthermore, also our test server is in a data center in Germany operated by the Hetzner Online GmbH.
The aim of this test setup is to be representative for a typical Internet connections in countries with a similar infrastructure like Germany.

To measure the RTTs between the involved peers, we require the IP address of each peer to conduct corresponding ping measurements.
While we have access to the IP address of our clients and the test server, we cannot look up the address of the client's locally configured DNS resolver.
Furthermore, the DNS resolvers might use an anycast service for its IP address~\cite{rfc1546} that may return different physical endpoints when pinged from the client and the server, respectively.
We used message~6 in Figure~\ref{fig:Recursive_DNS}, where the recursive resolver sends a request to the authoritative nameserver to learn the IP address of the recursive DNS resolver.
In detail, we announced a DNS authority section at our test server for a subdomain such as dnstest.example.com.
Then, we conducted a DNS query from the client to a random subdomain in our authority section such as foobar.dnstest.example.com.
At the same time, we captured the network traffic on the server and found a DNS query for this subdomain foobar.dnstest.example.com.
We reasoned that the sender's address of this DNS query is resolving the client's DNS query.
Depending on the DNS setup, the IP address of locally configured DNS resolver might mismatch the address sending the query to the authoritative nameserver.
For these cases, we assume that both DNS resolvers are colocated yielding about the same $\text{RTT}_{\text{DNS}}$ and $\text{RTT}_{\text{Server}}$ with respect to our measurements.

In total, we used 800  RIPE Atlas nodes in Germany to conduct our data collection on the 13th of June 2019.
A successful measurement includes $\text{RTT}_{\text{DNS}}$, $\text{RTT}_{\text{Server}}$, and $\text{RTT}_{\text{direct}}$ for the nodes, where we used an average over five ping measurements to determine the respective RTT.
In our data collection, we obtained successful results for 650 nodes.
Failures can be mainly attributed to DNS resolver that did not respond to ping measurements.
However, a small fraction of measurements experienced also failures during DNS measurements.

To focus our data collection on ISP-provided DNS resolvers, we investigated the autonomous system numbers of the observed IP addresses.
We assume, that an ISP-provided DNS resolver uses an IP address from the same autonomous system as the node does.
This approach allows us to sort out configured public DNS resolvers such as Google DNS which will usually operate from an IP address assigned to a different  autonomous system compared to the node.
In total, our data collection successfully obtained measurements from 474 nodes in Germany using each an ISP-provided DNS resolver.

\paragraph{Results}

\begin{figure}[t]
\centering
\includegraphics[width=0.47 \textwidth]{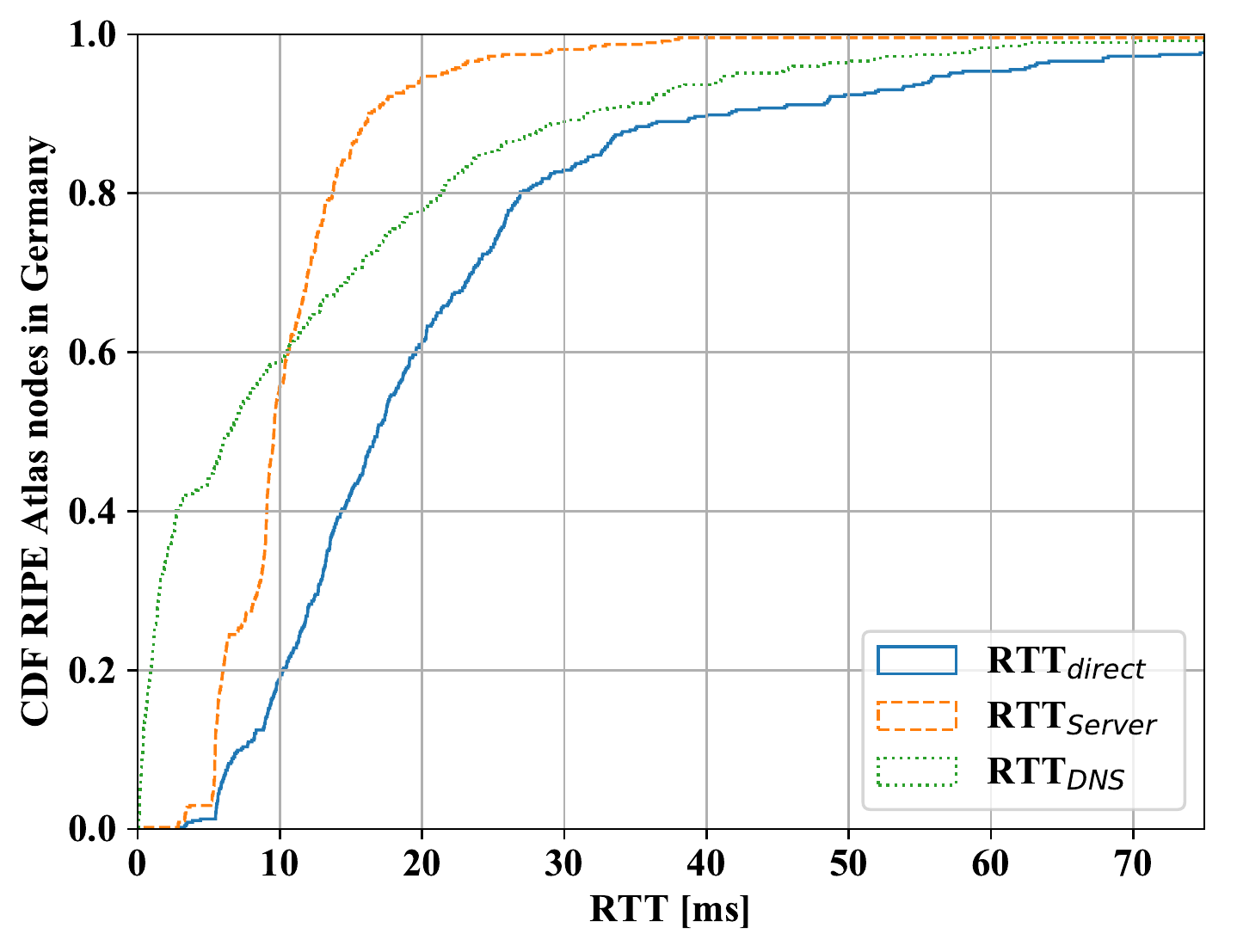}
\caption{Cumulative distribution of the RIPE Atlas nodes in Germany using an ISP-provided DNS resolver over $\text{RTT}_{\text{DNS}}$, $\text{RTT}_{\text{Server}}$, and $\text{RTT}_{\text{direct}}$.}
\label{fig:rtt_multi-isp}
\end{figure}

To accelerate a connection establishment via our proposal, we require $\text{RTT}_{\text{Server}}$ to be smaller than $\text{RTT}_{\text{direct}}$.
Figure~\ref{fig:rtt_multi-isp} provides a cumulative distribution of the RIPE Atlas nodes in Germany using an ISP-provided DNS resolver over the corresponding RTT's.
The plot shows $\text{RTT}_{\text{direct}}$, $\text{RTT}_{\text{Server}}$, and $\text{RTT}_{\text{DNS}}$ as solid, dashed, and dotted lines, respectively.
Our results indicate for almost all clients $\text{RTT}_{\text{Server}}$ is significantly smaller than $\text{RTT}_{\text{direct}}$.
For 51\% of the considered RIPE Atlas nodes, $\text{RTT}_{\text{Server}}$ is at least 5~ms smaller than $\text{RTT}_{\text{direct}}$.
Furthermore, 36.7\% of the nodes experience $\text{RTT}_{\text{Server}}$ to be at least 10~ms smaller than $\text{RTT}_{\text{direct}}$.
As can be observed in Figure~\ref{fig:rtt_multi-isp}, almost no nodes experiences $\text{RTT}_{\text{Server}}$ to be longer than 40~ms, while a tail of 10\% of the respective RIPE Atlas nodes observe a longer $\text{RTT}_{\text{direct}}$.
In this long tail, we find 7.2\% and 3.8\% of the nodes to have a $\text{RTT}_{\text{Server}}$ that outperforms $\text{RTT}_{\text{direct}}$ by at least 40~ms and 50~ms, respectively.
Furthermore, Figure~\ref{fig:rtt_multi-isp} provides a plot of $\text{RTT}_{\text{DNS}}$.
We find, that 60\% of the nodes have a RTT with their ISP-provided DNS resolver of less than 10~ms.
Moreover, $\text{RTT}_{\text{DNS}}$ is almost always smaller than $\text{RTT}_{\text{direct}}$ for a specific node.
This can be explained through RIPE Atlas nodes that are located towards the periphery of the Internet compared to their ISP-provided DNS resolvers holding a position closer to the core of the Internet.

\begin{figure}[t]
\centering
\includegraphics[width=0.47 \textwidth]{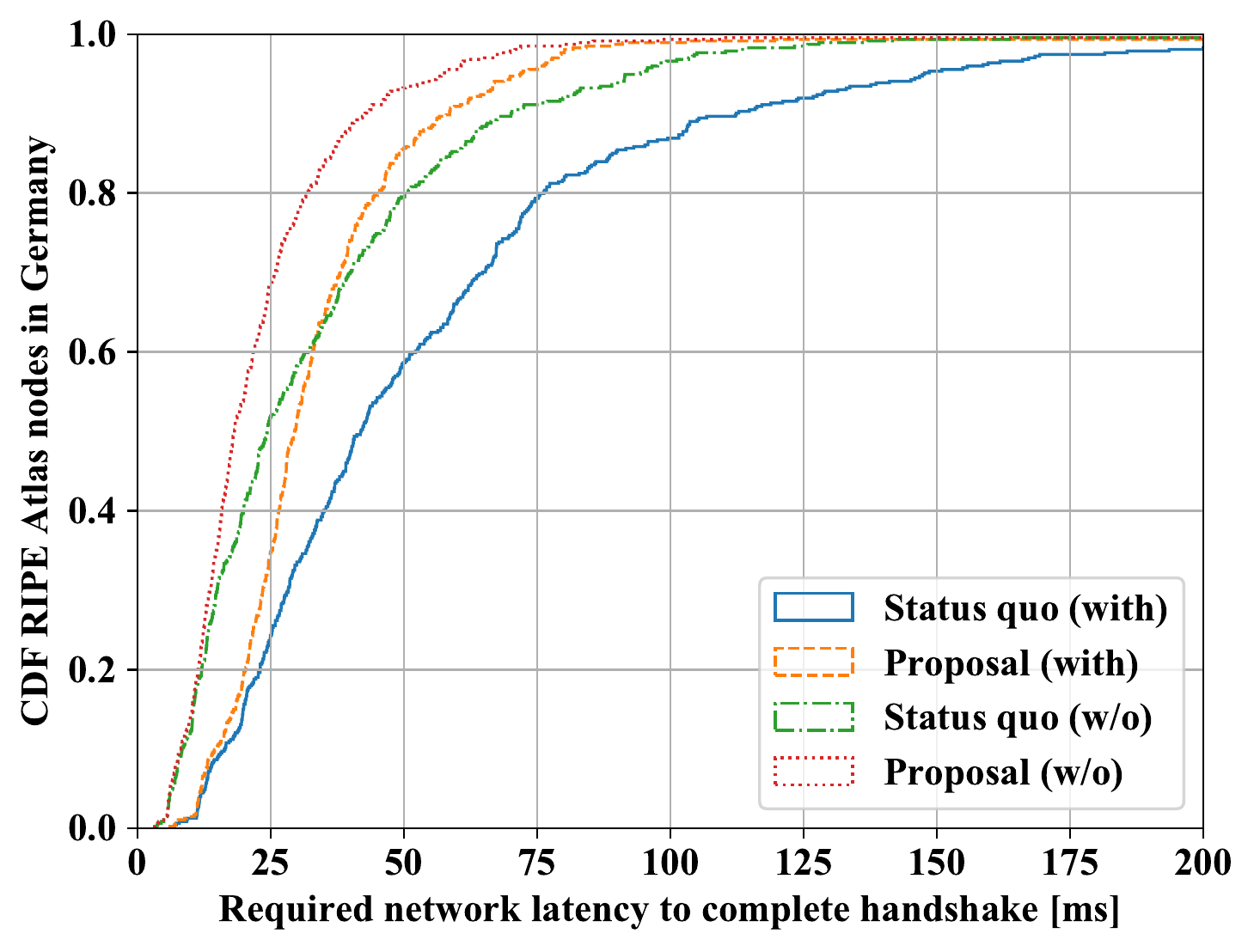}
\caption{Cumulative distribution of the RIPE Atlas nodes in Germany using an ISP-provided DNS resolver over the required network latency of the QUIC handshake. The plot compares the status quo versus our proposal for handshakes with and without stateless retry.}
\label{fig:default_versus_proposal-isp}
\end{figure}

To evaluate our proposal compared to the status quo, we combine the equations provided in Table~\ref{tab:analytical_model} with the measured RTT.
Figure~\ref{fig:default_versus_proposal-isp} plots these results as a cumulative distribution of the RIPE Atlas nodes in Germany using an ISP-provided DNS resolver over the required network latency to complete the QUIC connection establishment.
In total, Figure~\ref{fig:default_versus_proposal-isp} contains four plots.
In the scenario of a QUIC connection establishment using a stateless retry, the solid and dashed line represent the status quo and our proposed solution, respectively.
In the scenario of a QUIC handshake without stateless retry, the status quo and our proposal are marked as dash-dotted and dotted lines, respectively. 

In total, our results indicate our proposal accelerates the connection establishment for the great majority of investigated RIPE Atlas nodes.
Furthermore, we observe the trend that performance improvements are higher for nodes with longer required network latency to complete the handshake.
For example, we find that approximately 10\% of the nodes save at least 30~ms establishing the connection without stateless retry and 60~ms with a stateless retry.
Moreover, 24.3\% of the investigated nodes save at least 15~ms without and 30~ms with a stateless retry during the connection establishment.
Note, that approximately a third of the nodes experience a faster connection establishment using our proposal in a stateless retry connection establishment than having a status quo handshake without stateless retry. 

\subsection{Prototype-based Measurements}

In this section, we compare the delay of a default QUIC connection establishment with handshakes using our proposal.
However, the performance of our proposal significantly depends on the network topology of our test setup.
This measurement neglects network topologies and investigates the delay caused by the computational overhead of introducing a QuicSocks proxy on a network link.

\paragraph{Data collection}

For our test setup, we use a publicly accessibly QUIC server, a Dante SOCKS proxy (v1.4.2) and our implemented prototype to represent the client.
Our prototype and the Dante SOCKS proxy are run on the same virtual machine.
The virtual machine is equipped with 1~vCPU and 0.6~GB RAM and runs Debian 9.9 (Stretch).
The colocation of our client implementation with the proxy ensures that measurements using the proxy make use of the same network path as measurements conducted without the proxy.
In detail, we conduct three different types of measurements on the 25th of June 2019 of which we repeat each measurement 1\,000 times.
The \textit{default} measurements do not employ our proxy and investigate the required time to establish a QUIC connection with the server.
The \textit{cold start} measurements include the time required to establish the SOCKS connection and the subsequent QUIC handshake via the proxy.
Note, that a single SOCKS connection can be used to establish several QUIC connections. 
The \textit{warm start} measurement includes the time to establish a QUIC connection via our proxy but excludes the delay incurred by establishing the SOCKS connection.

\paragraph{Results}

Our data collection provided us with 1\,000 values for each of the three measurement types.
To evaluate our collected data, we retrieve the minimum and the median value of each measurement type. 
The \textit{default} measurement has a minimum of 49.145~ms and a median of 51.309~ms.
The \textit{warm start} measurement has a minimum of 49.708~ms and a median of 52.471~ms.
These values are between 1.1\% and 2.3\% higher than the \textit{default} measurement.
This can be explained by the additional overhead caused by the interaction with the proxy.
Furthermore, these values indicate an absolute overhead of using a SOCKS proxy of less than 1.2~ms for the median value, if the SOCKS connection is already established.
The \textit{cold start} measurement yields a minimum value of 52.073~ms and a median of 54.772~ms.
Comparing both measurements using the SOCKS proxy, we can attribute an additional overhead of about 2.3~ms in our test setup to establish the SOCKS connection.
As a result, we recommend clients to early establish their SOCKS connection and to use the \textit{warm start} approach to reduce the delays during their QUIC connection establishments.


%
%
%
%
%
%

\section{Related Work}\label{sec:Related}

There is much previous work on accelerating connection establishments on the web.
For example, Google launched in 2019 its feature Chrome Lite Pages~\cite{Lite_Pages}.
Lite Pages runs a proxy server that prefetches a website and forwards a compressed version of it to the client.
This approach leads to significant performance improvements for clients experiencing high network latencies as they do only need to establish a single connection to the proxy server to retrieve the website.
However, as major disadvantages compared to our proposal this leads to a significant load on the proxy server and breaks the principle of end-to-end transport encryption between the client and the web server.

Furthermore, Miniproxy~\cite{Siracusano:2016:FTA:2940147.2940149} can be used to accelerate TCP's connection establishment.
This approach places a proxy between the client and the web server, which doubles the number of required TCP handshakes.
Miniproxy can provide a faster TCP connection establishment in case of a favorable network topology and significant RTTs between client and web server.
The QUIC protocol includes computationally expensive cryptographic handshakes causing a significant delay compared to TCP's handshake~\cite{sy2019enhanced}.
Therefore, this approach seems less feasible when QUIC is used. 

The ASAP~\cite{Zhou:2011:ALT:2079296.2079316} protocol piggybacks the first transport packet within the client's DNS query and the DNS server forwards it to the web server after resolving the IP address.
However, this approach requires the DNS server to spoof the clients IP address which leads to a violation of the Best Current Practice RFC~2827~\cite{rfc2827}.
Furthermore, a deployment of ASAP requires significant infrastructural changes to the Internet because it uses a custom transport protocol.

Further possible performance improvements can be achieved by sending replicated DNS queries to several DNS resolvers and occasionally receiving a faster response~\cite{Vulimiri:2013:LLV:2535372.2535392}.
Another DNS-based mechanism aiming to reduce latency uses Server Push~\cite{rfc8484} where the resolver provides speculative DNS responses prior to the client's query.
In total, these approaches tradeoff a higher system utilization versus a possibly reduced latency.

\section{Conclusion} \label{sec:Conclusion}

We expect high-latency access networks to remain a web performance bottleneck for a significant number of users throughout the forthcoming years.
The QUIC protocols aims to reduce the delay of connection establishments on the web.
However, our measurements across a wide variety of access networks in Germany indicates a tail of users is affected by significant delays beyond 100~ms to complete a DNS lookup with a subsequent QUIC connection establishment.
Our proposal exploits the fact that ISP-provided DNS resolvers are typically located further into the core Internet than clients.
We find, that colocating a proxy with the ISP-provided DNS resolver provides significant performance gains for clients on high-latency access networks.
For example, a client can delegate the task of DNS lookups to the proxy in a more favorable network position.
Furthermore, the QUIC protocol provides features such as connection migration or the concept of stateless retries that allow further performance-optimizations when employing a proxy.
We hope that our work leads to an increased awareness of the performance problems experienced by a significant tail of users on high-latency access networks and spurs further research to reduce this web performance bottleneck.

\bibliographystyle{IEEEtran}
\bibliography{sample-bibliography}

\end{document}